\RequirePackage{lineno}
\documentclass[aps,floatfix,prl,twocolumn,showpacs,showkeys,amsmath,amssymb,superscriptaddress]{revtex4}

\usepackage{soul}
\usepackage{graphicx}	\usepackage{xspace}     \usepackage{amsmath}
\usepackage{hyperref}\usepackage{color}\usepackage{multirow}\usepackage{booktabs,array}

\makeatletter{}
\newcommand{\dhcs}{di-hadron correlations\xspace}

\newcommand{\dhc}{di-hadron correlation\xspace}
\newcommand{\pT}{$p_{\rm T}$\xspace}
\newcommand{\IAA}{$I_{AA}$\xspace}

\newcommand{\dphi}{$\Delta\phi$\xspace}

\newcommand{\deta}{$\Delta\eta$\xspace}
\newcommand{\GeV}{GeV/$c$\xspace}

\newcommand{\vn}{$v_{n}$\xspace}
\newcommand{\vnt}{$v_{n}^{t}$\xspace}
\newcommand{\vneff}{$\tilde{v}_{n}^{t}$\xspace}
\newcommand{\vneffnot}{$\tilde{v}_{n}$\xspace}
\newcommand{\vnaeff}{$\tilde{v}_{n}^{a}$\xspace}
\newcommand{\vnteff}{$\tilde{v}_{n}^{t}$\xspace}

\newcommand{\vnumeffassoc}[1]{$\tilde{v}_{#1}^{a}$\xspace}
\newcommand{\vnumefftrigger}[1]{$\tilde{v}_{#1}^{t}$\xspace}

\newcommand{\vnefftrigger}{$\tilde{v}_{n}^{\mathrm{t}}$\xspace}
\newcommand{\vneffassoc}{$\tilde{v}_{n}^{\mathrm{a}}$\xspace}

\newcommand{\eref}[1]{equation~\ref{#1}}

\newcommand{\Fref}[1]{Figure~\ref{#1}}
\newcommand{\Tref}[1]{Table~\ref{#1}}
\newcommand{\Eref}[1]{Equation~\ref{#1}}

\newcommand{\pp}{$p$+$p$\xspace}

\newcommand{\Au}{$Au+Au$\xspace}
\newcommand{\Pb}{$Pb+Pb$\xspace}
\newcommand{\dAu}{$d+Au$\xspace}

\newcommand{\AplusA}{$A$+$A$\xspace}

\newcommand{\sNN}{$\sqrt{s_{\mathrm{NN}}}$}
\newcommand{\ns}{near-side\xspace}
\newcommand{\as}{away-side\xspace}

\newcommand{\nas}{near- and away-side\xspace}

\newcommand{\ptassoc}{$p_T^{\mathrm{a}}$\xspace}

\newcommand{\psisrange}[2]{#1 $<\phi_s<$ #2\xspace}
\newcommand{\pttrigrange}[2]{#1 $< p_T^{\mathrm{t}} <$ #2~GeV/$c$\xspace}
\newcommand{\ptassocrange}[2]{#1 $< p_T^{\mathrm{a}} < $ #2~GeV/$c$\xspace}

\newcommand{\inplane}{in-plane\xspace}
\newcommand{\outplane}{out-of-plane\xspace}

\begin{document}

\title{Disappearance of the Mach Cone in heavy ion collisions}

\author{Christine Nattrass} 
\affiliation{University of Tennessee, Knoxville, TN, USA-37996.}
\author{Natasha Sharma} \affiliation{University of Tennessee, Knoxville, TN, USA-37996.}
\affiliation{Department of Physics, Panjab University, Chandigarh, India-160014.}
\author{Joel Mazer}
\author{Meghan Stuart}
\author{Aram Bejnood}
\affiliation{University of Tennessee, Knoxville, TN, USA-37996.}
\date{\today}

\begin{abstract} 
\makeatletter{}We present an analysis of \dhcs using recently developed methods for background subtraction which allow for higher precision measurements with fewer assumptions about the background.  These studies indicate that low momentum jets interacting with the medium do not equilibrate with the medium, but rather that interactions with the medium lead to more subtle increases in their widths and fragmentation functions, consistent with observations from studies of higher momentum fully reconstructed jets.  The \as shape is not consistent with a Mach cone.  The qualitatively different conclusions reached with a more careful consideration of the background subtraction call into question the complete suppression of jets in central collisions observed in earlier studies, indicating that this is also an artifact of the background subtraction. 

\end{abstract}

  \maketitle

 \section{Introduction}\label{Sec:Introduction}
\makeatletter{}A hot, dense medium called the Quark Gluon Plasma (QGP) is created in high energy heavy ion collisions~\cite{Adcox:2004mh,Adams:2005dq,Back:2004je,Arsene:2004fa}.  Quarks and gluons scattered early in the collision live through the QGP phase and interact strongly with the medium.  This leads to the suppression of hadrons at high transverse momenta relative to expectations from binary scaling in \pp collisions~\cite{Adams:2003kv,Adler:2003qi,Back:2004bq,Aamodt:2010jd,CMS:2012aa}, a process called jet quenching.
One of the early indications of jet quenching was measurements of \dhcs at high transverse momenta (\pT)~\cite{Adler:2002tq,Adler:2005ee}.  A high momentum particle created in an \AplusA collision is used as a proxy for a jet, assuming that a sufficiently high \pT particle originated from a jet and that its direction is approximately the direction of the parton.  The distribution of softer particles in the event, called associated particles, can be measured in azimuth (\dphi) and pseudorapidity (\deta) relative to the trigger particle.  The correlation function has two peaks in azimuth, one near the trigger particle ($\Delta\phi\approx0$), called the \ns, and one 180$^\circ$ away, called the \as.  The \ns is roughly comparable to that observed in \dAu and \pp collisions, with slight modifications~\cite{Agakishiev:2011st,Abelev:2009ah,Abelev:2009af}.  Numerous studies reported modifications of the \as shape, observing a local minimum rather than a peak~\cite{Adler:2005ee,Adare:2006nr,Adare:2007vu,Abelev:2008ac,Aggarwal:2010rf}.  This was frequently interpreted as a Mach cone from a parton propagating faster than the speed of sound in the QGP~\cite{Torrieri:2009mv}.  We refer to this as the Mach cone below.  This feature indicated a qualitatively different interaction with the medium for the soft jets which dominate \dhcs and the hard jets observed at the Large Hadron Collider (LHC).  Studies at the LHC indicated that the fragmentation functions were slightly modified~\cite{Aad:2014wha,Chatrchyan:2014ava} but studies of fully reconstructed jets have not indicated any severe shape modifications.

The combinatorial background in azimuth in \dhcs has the form
\begin{equation}
 \frac{dN}{\pi d\Delta\phi} = B( 1 + \sum_{n=2}^{\infty} 2 \tilde{v}_{n}^{\mathrm{t}} \tilde{v}_{n}^{\mathrm{a}} \cos(n\Delta\phi))\label{Eqtn:JBBBCorrelations}
\end{equation}
\noindent where \vnefftrigger (\vneffassoc) are the Fourier coefficients of the trigger (associated) particles in the background~\cite{Bielcikova:2003ku}.  The \vneffnot may arise due to hydrodynamical flow or jet quenching.
The majority of \dhc studies used the zero yield at minimum (ZYAM) method~\cite{Adams:2005ph,PunchThrough,Adare:2008ae,Ajitanand:2005jj}, or some variation thereof, to determine the background level $B$, combined with the assumption that independent measurements of the \vn were the appropriate \vneffnot for these studies.  The odd \vn were assumed to be zero until it was proposed that non-zero odd \vn could arise from fluctuations in the initial condition~\cite{Sorensen:2010zq,Alver:2010gr}.  Non-zero odd \vn were subsequently measured ~\cite{ALICE:2011ab,Adamczyk:2013waa,Adare:2011tg}.  A majority of the published \dhcs therefore neglect the odd \vn.
While the observation of odd \vn indicated that at least the magnitude of the Mach Cone signal in the data was overestimated, theoretical studies of the Mach Cone continue~\cite{Bouras:2012sx,Bouras:2012mh,Ayala:2012bv,Bouras:2014rea,Tachibana:2015qxa}.  The few \dhc studies since the observation of odd \vn are either inconclusive about the presence or absence of shape modifications~\cite{Adare:2012qi} or indicate that the shape modification persists~\cite{Agakishiev:2014ada}.

We reassess the data using improved background subtraction techniques in order to determine the presence or absence of shape modifications and whether or not \dhcs are consistent with studies using fully reconstructed jets.  In~\cite{Sharma:2015qra} we presented an alternate method for determining the background which overcomes many of the limitations of ZYAM, making fewer assumptions about the shape and level of the background.  
The Reaction Plane Fit (RPF) method uses the correlation functions in the background dominated region on the \ns ($\Delta\eta>0.7$, $\Delta\phi<\pi/2$) and the fact that the \vneff depend on the reaction plane 
when the angle of the trigger particle is restricted relative to the event plane~\cite{Bielcikova:2003ku}.  The correlation functions in bins of the angle of the trigger particle relative to the event plane are fit simultaneously to determine the $B$, \vnaeff, and \vneff in \eref{Eqtn:JBBBCorrelations}.  
This method makes fewer assumptions about the shape of the background than other methods.
In this paper we apply this method to \dhcs before background subtraction measured by STAR~\cite{Agakishiev:2010ur,Agakishiev:2014ada}.  We briefly review the relevant details of the measurement and then present \dhcs using the RPF method~\cite{Sharma:2015qra} for background subtraction. 

\section{Analysis}\label{testthis}\makeatletter{}In~\cite{Agakishiev:2010ur} correlation functions were reported for \dAu collisions at \sNN = 200 GeV and for 20--60\% \Au collisions at \sNN = 200 GeV in bins of the angle between the trigger particle and the reconstructed reaction plane, $\phi_s = \phi^t - \psi$ in two regions in \deta ($|\Delta\eta|<$~0.7 and 0.7~$<|\Delta\eta|<$~2.0).  
The correlation functions were normalized by the number of trigger particles rather than by the number of events so all bins in $\phi_s$ have the same $B$ in \eref{Eqtn:JBBBCorrelations}.  The \vneff in a given $\phi_s$ bin depend on the range of the bin in $\phi_s$, the \vnt,  and the reaction plane resolution~\cite{Bielcikova:2003ku}.  The systematic uncertainties on the reaction plane resolution were 1\% for the second order reaction plane and 3\% for the fourth and sixth~\cite{Agakishiev:2010ur}.  

The background is fit up to n~=~4 in \eref{Eqtn:JBBBCorrelations}.  The uncertainty due to the reaction plane resolution uncertainty is determined by varying the reaction plane resolution and refitting the background.
The background determined from the background dominated region (0.7~$<|\Delta\eta|<$~2.0) must be scaled to determine the appropriate background level in the signal+background region ($|\Delta\eta|<$~0.7).  The correlation functions in~\cite{Agakishiev:2010ur} were not corrected for acceptance effects using mixed events, as in~\cite{Abelev:2009af,Agakishiev:2011st,Sharma:2015qra}.  If the distribution of particles in the event were independent of $\eta$, the exact scaling factor could be determined analytically, however, the $\eta$-dependent track distribution and tracking efficiency alter the ratio of the number of track pairs in the background-dominated region to the signal-dominated region.  The background is scaled up by 2\% more than expectations from an 
$\eta$-independent track distribution with a systematic error of 1\%.  The scale for the background is determined by comparison to the highest momentum associated momentum (\ptassocrange{3.0}{4.0}) and trigger momentum (\pttrigrange{4.0}{6.0}).  The same scaling is used for all momenta.  Note that the scale uncertainty is not a feature of the method and would be avoidable in future studies.  The scale and reaction plane uncertainties are correlated point-to-point and when comparing correlations at different momenta.   The statistical uncertainties on the \Au data include uncertainties due to the background determined by the fit in the RPF method and are therefore non-trivially correlated point-to-point.
Since the RPF method assumes that the residual signal in the background-dominated region is negligible, we restrict our study to high momenta (\pttrigrange{4.0}{6.0} and \ptassocrange{1.5}{4.0}) where PYTHIA~\cite{Sjostrand:2006za,Skands:2010ak} studies indicate that the \nas peaks are well separated.   The \dAu data in~\cite{Agakishiev:2010ur} are background subtracted and are therefore compared to the \Au data without modification.

\begin{figure*}
\begin{center}
\rotatebox{0}{\resizebox{15cm}{!}{
        \includegraphics{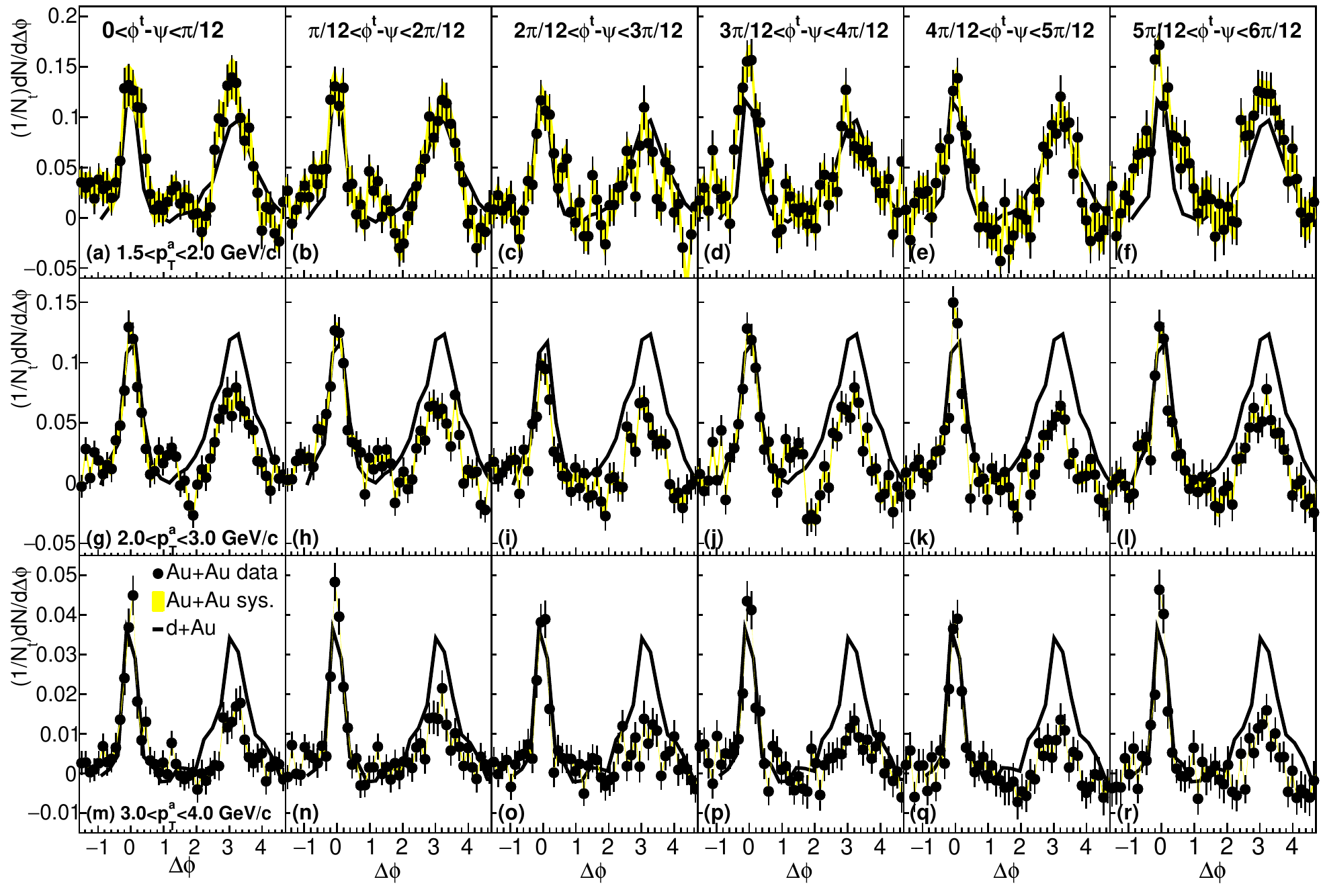}
}}
\caption{
Background subtracted \dhcs with \pttrigrange{4.0}{6.0} for \ptassocrange{1.5}{2.0} (a-f), \ptassocrange{2.0}{3.0} (g-l), \ptassocrange{3.0}{4.0} (m-r) in \dAu and \Au collisions at \sNN = 200 GeV for trigger particles from \psisrange{0}{$\pi/12$}(a,g,m), \psisrange{$\pi/12$}{$2\pi/12$}(b,h,n), \psisrange{$2\pi/12$}{$3\pi/12$}(c,i,o), \psisrange{$3\pi/12$}{$4\pi/12$}(d,j,p), \psisrange{$4\pi/12$}{$5\pi/12$}(e,k,q), and \psisrange{$5\pi/12$}{$6\pi/12$}(f,l,r).  Yellow band shows scale uncertainty on \Au data.  Uncertainty due to reaction plane resolution is not shown but is negligible.  The statistical uncertainties on the \Au data include uncertainties due to the background determined by the fit in the RPF method and are therefore non-trivially correlated point-to-point.}\label{Fig:Correlations}
\vspace{-0.20cm}
\end{center}
\end{figure*}

\begin{table*}
\begin{center}
\caption{Fit parameters from RPF method.  The $v_{2}$(\%) from~\cite{Agakishiev:2010ur} are listed for comparison to \vnumeffassoc{2}(\%).  For 4.0~$<$~\pT~$<$6.0 \GeV $v_{2}$~=~16.3~$\pm$~2.0 from~\cite{Agakishiev:2010ur}.}
\label{Tab:vn}
\begin{tabular}{c | c c c c c c c  c}
\ptassoc (\GeV) & $\chi^2$/NDF & B & $v_{2}$(\%) from~\cite{Agakishiev:2010ur} &   \vnumeffassoc{2}(\%) & \vnumefftrigger{2}(\%) & $\tilde{v}_{3}^{t}\tilde{v}_{3}^{a}$ ($\times$10$^{-4}$) &   \vnumeffassoc{4}(\%) & \vnumefftrigger{4}(\%)\\ \hline
1.5--2.0 & 1.00 & 0.7 $\pm$ 0.0 & 16.4 $\pm$ 1.1 & 18.1 $\pm$ 0.2 & 9.3 $\pm$ 1.6 & 109 $\pm$ 17 & 2.6 $\pm$ 0.4 & 0.10 $\pm$ 0.08 \\ 
2.0--3.0 & 1.23 & 0.3 $\pm$ 0.0 & 18.9 $\pm$ 1.2 & 20.7 $\pm$ 0.3 & 8.7 $\pm$ 2.4 & 159 $\pm$ 28 & 4.4 $\pm$ 0.7 & 0.10 $\pm$ 0.08 \\ 
3.0--4.0 & 1.04 & 0.0 $\pm$ 0.0 & 19.4 $\pm$ 1.3 & 22.9 $\pm$ 1.0 & 10.8 $\pm$ 7.0 & 190 $\pm$ 89 & 2.7 $\pm$ 2.1 & 0.10 $\pm$ 0.07 \\

\end{tabular}
\end{center}
\end{table*}

Fit parameters from the RPF method are given in \Tref{Tab:vn}.  The \vnteff are within error for all \ptassoc and the \vnumefftrigger{4} are consistent with zero.  The \vnumeffassoc{2} (\vnumefftrigger{2}) determined from the fit are larger (smaller) than the $v_{2}$ from inclusive particles from~\cite{Agakishiev:2010ur}.  There are several possible explanations for differences between the inclusive \vn and the \vneff determined from the RPF method.  The terms \Eref{Eqtn:JBBBCorrelations} should be the weighted average of the product $ \tilde{v}_{n}^{t}\tilde{v}_{n}^{a} = \langle v_{n}^{t}v_{n}^{a} \rangle$, rather than $\langle v_{n}^{t}\rangle\langle v_{n}^{a} \rangle$, and the average should be over background pairs.  The centrality bin is wide and high \pT trigger particles are more likely to be in more central events, meaning that the background pair-averaged \vneff in~\Eref{Eqtn:JBBBCorrelations} is not necessarily the same as the inclusive \vn.  Events which contain high \pT particles may not have the same properties as inclusive events, or the presence of a high \pT particle may be indicative of a hot spot in the medium.  Imperfect correlations between the n~=~2 and higher order reaction planes~\cite{Aad:2014fla}, possibly due to a difference between the reaction plane for flow and jet quenching~\cite{Jia:2012ez}, would reduce the \vnteff and increase \vnaeff but the RPF method would still provide a valid description of the background.  The fit parameters in \Tref{Tab:vn} therefore should not be considered a measurement of the \vn due to flow.  While the RPF method assumes that the large \deta correlation function on the \ns contains no signal, no assumptions are made about the event sample and the background is weighted correctly over background pairs by construction.  The background determined using the RPF method is therefore more robust than that determined using the ZYAM method.

Background subtracted \dhcs with \pttrigrange{4.0}{6.0} and \ptassocrange{1.5}{4.0} in \dAu and \Au collisions at \sNN = 200 GeV are shown in \Fref{Fig:Correlations}. 
Typical uncertainties in \Fref{Fig:Correlations} are less than the uncertainty due to the background level alone using the ZYAM method and the uncertainty in the \ns \inplane peak region is about one fifth of the uncertainty using ZYAM~\cite{Agakishiev:2010ur}.
The data are inconsistent with the presence of the Mach cone~\cite{Adler:2005ee,Adare:2006nr,Adare:2007vu,Abelev:2008ac,Aggarwal:2010rf}, indicating that it was an artifact of the background subtraction.  

\begin{figure*}
\begin{center}
\rotatebox{0}{\resizebox{15cm}{!}{
        \includegraphics{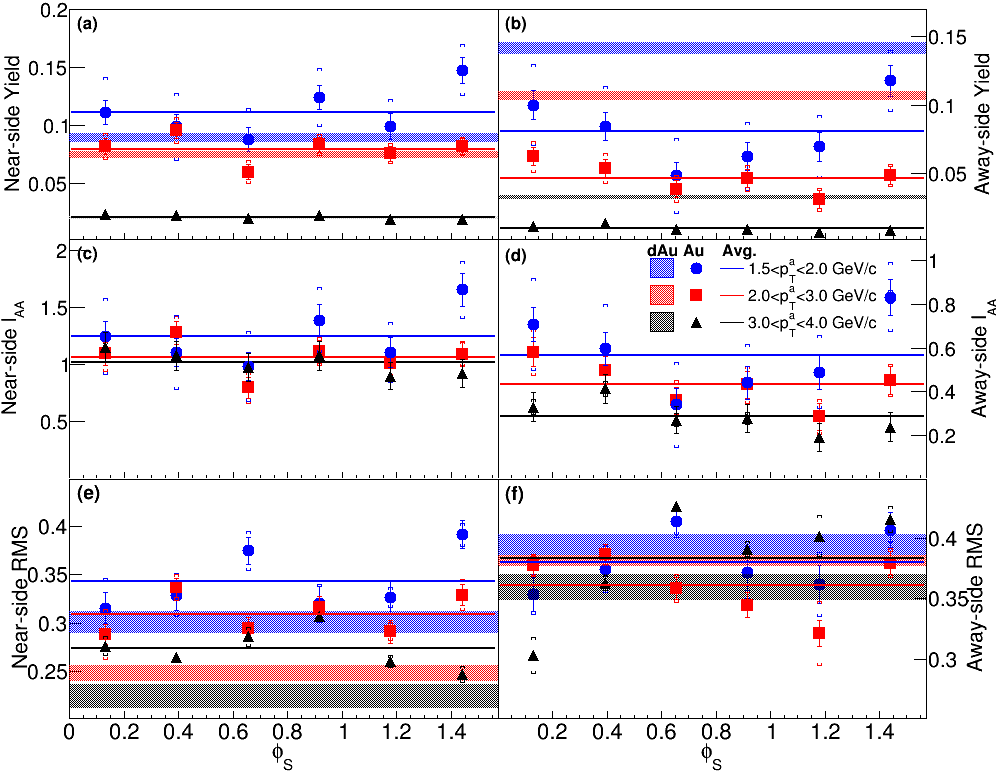}
}}
\caption{
Yields (a,b), truncated RMS (c,d), and \IAA (e,f) for the \ns (a,c,e) and \as (b,d,f).  Systematic uncertainties are 100\% correlated point to point.  Statistical uncertainties are non-trivially correlated point to point for a fixed \ptassoc but are uncorrelated for different \ptassoc.  Lines show the average for all $\phi_s$ and bands show the value for \dAu collisions with the width representing the error bar.}\label{Fig:YieldAndRMS}
\vspace{-0.20cm}
\end{center}
\end{figure*}

The yield is reported in order to quantitatively compare correlation functions in \Au collisions to \dAu collisions.  The yield is calculated as
\begin{equation}
 Y = \int_{a}^{b}  \frac{1}{N_{t}}\frac{dN}{d\Delta\phi}  d\Delta\phi\label{Eqtn:Yield}
\end{equation}
\noindent where the integration limits $a=-0.79$ and $b=0.79$ on the \ns and $a=2.36$ and $b=3.93$ on the \as are chosen to match the binning in the STAR data.  
The \ns yield is given in \Fref{Fig:YieldAndRMS}(a) and the \as yield in \Fref{Fig:YieldAndRMS}(b).  These results have substantially smaller uncertainties than results using the ZYAM method~\cite{Adare:2010mq}.  The yield is highest for the lowest momenta.   The \ns yields in \Au collisions are within error of the yields in \dAu collisions but the \as yields are below those in \dAu collisions.

\IAA has been used to quantify the suppression observed in \dhcs~\cite{Adare:2008ae}, analogous to measurements of the nuclear modification factor.  We calculate \IAA as
\begin{equation}
 I_{AA} = Y_{Au+Au}/Y_{d+Au}.\label{Eqtn:IAA}
\end{equation}
\noindent   Note that \eref{Eqtn:IAA} differs from the standard definition of \IAA, which has the yield from \pp collisions in the denominator.   No differences have been observed between di-hadron correlations in \pp and \dAu~\cite{Adler:2005ad} and only \dAu data are available in~\cite{Agakishiev:2010ur}.
The \ns \IAA is given in \Fref{Fig:YieldAndRMS}(c) and the \as \IAA in \Fref{Fig:YieldAndRMS}(d).  The \ns \IAA is within error of 1.0 with little reaction plane dependence, although \IAA is also consistent with the slight enhancement observed at the LHC~\cite{Aamodt:2011vg}.  The \as \IAA is around 0.3 for \ptassocrange{3.0}{4.0}, increasing with increasing \ptassoc.  The higher \IAA for lower \ptassoc is consistent with the softening of the fragmentation function expected in response to the medium.  \IAA is highest both in- and out-of-plane for \ptassocrange{1.5}{2.0}.  A similar trend is indicated for \ptassocrange{2.0}{3.0}, although with less significance.  Correlations with a trigger \outplane may be more surface biased than those \inplane and therefore less modified, or they may interact with more medium.  The reaction plane dependence observed may be due to an interplay between these effects.  

Early results demonstrated a complete suppression of the \as in central collisions~\cite{Adler:2002tq}, a highly cited and influential result.  The \IAA in~\cite{Adler:2002tq} is consistent within the large uncertainties with \Fref{Fig:YieldAndRMS}(a) and \Fref{Fig:YieldAndRMS}(b) for the same centralities.  However, since we observe qualitatively different results from studies using the same methodology~\cite{Adler:2005ee,Adare:2006nr,Adare:2007vu,Abelev:2008ac,Aggarwal:2010rf}, the complete suppression of the \as in~\cite{Adler:2002tq} is likely also an artifact of the background subtraction.

We also report the truncated RMS
\begin{equation}
 RMS = \sqrt{ \int_{a}^{b}  \frac{1}{Y N_{t}}\frac{dN}{d\Delta\phi} (\Delta\phi - \Delta\phi_0)^2 d\Delta\phi }\label{Eqtn:RMS}
\end{equation}
\noindent where $\Delta\phi_0 = 0$ for the \ns and $\pi$ for the \as rather than the RMS over all \dphi because integration over a wide range in \dphi increases the weight of statistical error bars without dramatically changing the result.  The integration limits are the same as for \eref{Eqtn:Yield}.
The \ns RMS is given in \Fref{Fig:YieldAndRMS}(e) and the \as RMS in \Fref{Fig:YieldAndRMS}(f).   The \ns RMS is larger than the \dAu RMS for the same \ptassoc for all ranges of \ptassoc studied.  Such broadening is consistent with expectations from energy loss through either bremsstrahlung or collisional energy loss, since the energy would remain spatially correlated with the parent parton but be distributed over a somewhat wider area~\cite{Wiedemann:2009sh}.  This indicates that the \ns is modified even though the yields are consistent with those in \dAu collisions, consistent with observations in~\cite{Agakishiev:2011st}.  There is little indication of $\phi_s$-dependent modifications on the \ns.
The \as RMS in \Au collisions is mostly consistent with that in \dAu collisions.  Since broadening is apparent on the \ns, this may indicate that the \as RMS is less sensitive to observation of broadening because the \as is already broader than the \ns, even in \dAu collisions, due to the difference between the trigger particle's angle and the angle of the \as jet.

These high precision results are consistent with the modification of fragmentation functions in \Pb collisions observed at the LHC, which indicated broadening and softening of the fragmentation function~\cite{Aad:2014wha,Chatrchyan:2014ava}.  This demonstrates the efficacy of the RPF method for precision studies of di-hadron and jet-hadron correlations, particularly at low momenta.

\section{Conclusions}\label{Sec:Conclusions}
 \makeatletter{}The effects observed in \dhcs relative to the reaction plane using the RPF method indicate that medium-induced modifications to jet-like correlations are more subtle than earlier results using the ZYAM method and neglecting odd \vn.  This indicates that jets do not equilibrate fully with the medium but lost energy remains spatially correlated with the parent parton and that the complete suppression observed in~\cite{Adler:2002tq} was likely also an artifact of the background subtraction.  These results agree better with results from fully reconstructed jets, which do not indicate dramatic shape modifications or complete equilibration but slight broadening and more subtle modifications of the fragmentation function~\cite{Aad:2014wha,Chatrchyan:2014ava}, than with earlier results indicating a dip~\cite{Adler:2005ee,Adare:2006nr,Adare:2007vu,Abelev:2008ac,Aggarwal:2010rf}, or ``Mach cone``, on the \as. 
 \section{Acknowledgements}
 \makeatletter{}We are grateful to Jana Biel\v{c}ikova, Marco van Leeuwen, Soren Sorensen, and Paul Stankus for useful comments on the manuscript and to Frank Geurts, Fuqiang Wang, and the STAR collaboration for productive communication about the STAR data.  This work was supported in part by funding from the Division of Nuclear Physics of the U.S. Department of Energy under Grant No. DE-FG02-96ER40982. N.S. acknowledges the support of 268 DST-SERB Ramanujan Fellowship (D.O. No. SB/S2/RJN- 269084/2015). 

\makeatletter{}
    
\end{document}